\documentclass[11pt]{article}
\usepackage{stmaryrd}
\usepackage{bbm}
\usepackage{mathrsfs}
\usepackage{amsfonts,color}
\usepackage{amssymb}
\usepackage{amsmath}
\usepackage{amsthm}
\usepackage{amsmath}
\usepackage{accents}
\usepackage{lscape,cite}
\usepackage[figuresright]{rotating}
\usepackage{graphicx}
\usepackage{animate}

\newcommand{\al}{\alpha}
\newcommand{\be}{\beta}
\newcommand{\ga}{\gamma}
\newcommand{\de}{\delta}

\def \h#1{\widehat{#1}}
\def \t#1{\widetilde{#1}}
\def \b#1{\overline{#1}}

\def \th#1{\widehat{\widetilde{#1}}}

\makeatletter
\def\wideubar{\underaccent{{\cc@style\underline{\mskip20mu}}}}
\def\Wideubar{\underaccent{{\cc@style\underline{\mskip8mu}}}}
\makeatother
\makeatletter
\def\wideubar{\underaccent{{\cc@style\underline{\mskip8mu}}}}
\def\Wideubar{\underaccent{{\cc@style\underline{\mskip6mu}}}}
\def\wideutilde{\underaccent{{\cc@style\undertilde{\mskip6mu}}}}
\def\ut{\underaccent{{\cc@style\undertilde{\mskip6mu}}}}
\def\ubb{{\underaccent{{\cc@style\underline{\mskip6mu}}}}}
\makeatother

\makeatletter
\def\widebar{\accentset{{\cc@style\underline{\mskip10mu}}}}
\def\Widebar{\accentset{{\cc@style\underline{\mskip8mu}}}}
\makeatother

\allowdisplaybreaks[4]
\begin{document}

\title{Addition formulae, B\"acklund transformations, periodic solutions and quadrilateral equations }

\author{Danda Zhang$^{1}$,~ Da-jun Zhang$^{2}$\footnote{Corresponding author. Email: djzhang@staff.shu.edu.cn}\\
{\small $^{1}$Faculty of Science, Ningbo University, Ningbo 315211, P.R. China}\\
{\small $^{2}$Department of Mathematics, Shanghai University, Shanghai 200444, P.R. China}}

\date{\today}
\maketitle

%---------------------------------------------------------------------------------------------
%---------------------------------------------------------------------------------------------

\begin{abstract}

Addition formulae of trigonometric and elliptic functions are used to generate
B\"acklund transformations together with their connecting quadrilateral equations.
As a result we obtain periodic solutions for a number of multidimensionally consistent
affine linear and multiquadratic quadrilateral equations.

\vspace {10pt}
\noindent
{\bf Keywords:} addition formulae, trigonometric functions, elliptic functions, B\"{a}cklund transformation, quadrilateral equations
\\
{\bf PACS numbers:}  02.30.Ik, 05.45.Yv

\end{abstract}

\section{Introduction}\label{sec-1}

There has been a surge of interest in discrete integrable systems in the last two decades (cf.\cite{HJN-book-2016} and the references therein).
Among variety of discrete integrable systems, quadrilateral equations,
equipped with multidimensional consistency \cite{NijW-GMJ-2001,Nij-PLA-2002,BS-IMRN-2002},
provide one of the simplest types of integrable partial difference equations
and are playing important roles in the research of  discrete integrable systems.
Scaler multidimensional consistent quadrilateral equations were classified with extra restrictions (affine linearity, $D4$ symmetry and
tetrahedron property) and the results are known as the Alder-Bobenko-Suris (ABS) list \cite{ABS-CMP-2003}
in which equations are named as
Q4, Q3($\delta$), Q2,  Q1($\delta$), A2, A1($\delta$),  H3($\delta$), H2 and H1,
where $\delta$ is a parameter. These equations have been extended to multiquadratic  case \cite{AtkN-IMRN-2014}
in which the equations are labeled by  Q4$^{*}$, Q3$^{*}$($\delta$), etc.

Generic form of a quadrilateral equation is
\begin{equation}
Q(u, \t u, \h u, \th u; p,q)=0,
\label{Q}
\end{equation}
where
\begin{equation}
 u\doteq u(n,m) \doteq u_{n,m},~ \t u \doteq u_{n+1,m}, ~\h u \doteq u_{n,m+1},~\th u \doteq u_{n+1,m+1},
\end{equation}
and $p$ and $q$ are respectively spacing parameters of $n$ and $m$ direction.
If Eq.\eqref{Q} is multidimensional consistent, one can immediately write out its B\"acklund transformation (BT) of the following form
\begin{subequations}\label{BT-Q}
\begin{align}
& Q(u, \t u, \b u, \t{\b u}; p,r)=0,\\
& Q(u, \b u, \h u, \h{\b u}; r,q)=0,
\end{align}
\end{subequations}
where $\b u$ is viewed as a new solutions of Eq.\eqref{Q}
as well as a shift of $u$ along the third direction $l$ (e.g. $\b u \doteq u_{n,m,l+1}$) and $r$ is the spacing parameter of $l$ direction.
Such BTs are useful in generating solutions to Eq.\eqref{Q} (cf.\cite{HieZ-JPA-2009,ZhaZ-SIGMA-2017,ZhaZ-JNMP-2018}).

Addition formulae of periodic functions such as trigonometric functions and elliptic functions
can be considered as one-step shift relations. For example,
\[\sin(\al- \be)=\sin(\al)\cos(\be)- \cos(\al)\sin(\be)\]
indicates a shift relation $u\t U-\t u U=p$ if taking $u=\cos(\alpha)$, $U=\sin(\al)$, $p=\sin(a)$,
$\al=an+bm$ and $\beta=\t \al$.
Generic form of the above shift relation is
\begin{subequations}\label{BT-f}
\begin{equation}
f(u,U,\t u, \t U,p)=0,
\end{equation}
and symmetrically, we have
\begin{equation}
f(u,U,\h u, \h U,q)=0.
\end{equation}
\end{subequations}
These two equations are considered as a solvable coupled system. Then compatibility is required.
In principle, the  compatibility w.r.t. $U$ (i.e. $\th U=\t{\h U}$) leads to  a quadrilateral equation of $u$,
\begin{equation}
F(u,\t u, \h u, \th u; p,q)=0,
\label{u}
\end{equation}
while the  compatibility w.r.t. $u$ yields a quadrilateral equation of $U$,
\begin{equation}
G(U,\t U, \h U, \th U; p,q)=0.
\label{U}
\end{equation}
Eqs.\eqref{BT-f}, \eqref{u} and \eqref{U} compose a consistent triplet,
in which \eqref{BT-f} acts as a BT to connect \eqref{u} and \eqref{U}.
Such triplets have been used as a main tool to construct rational solutions for a number of
quadrilateral equations \cite{ZhaZ-SIGMA-2017}.

In this paper, inspired by some addition formulae of trigonometric functions and elliptic functions,
we construct systems of the coupled form \eqref{BT-f}.
Then, for each system by checking its compatibility we find a consistent triplet that consists of two quadrilateral equations and
their BT in the form \eqref{BT-f}.
Meanwhile, we obtain periodic solutions of the  consistent triplet.
All the results we obtain in the paper are listed out at the end of the paper.

The paper is organized as follows.
In Sec.\ref{sec-2} addition formulae of trigonometric functions are used to generate  consistent triplets and their
period solutions in terms of  trigonometric functions.
Then, in a similar manner in Sec.\ref{sec-3} we  obtain elliptic function solutions for some quadrilateral equations
by using addition formulae of elliptic functions.
Finally Sec.\ref{sec-4} consists of conclusion and discussion.

\section{Cases of trigonometric functions}\label{sec-2}

Addition formulae of trigonometric functions of our interests in the paper are the following,
\begin{subequations}\label{trig-id}
\begin{align}
&\sin(\al\pm \be)=\sin(\al)\cos(\be)\pm \cos(\al)\sin(\be),\label{trig-id1}\\
&\cos(\al\pm \be)=\cos(\al)\cos(\be)\mp \sin(\al)\sin(\be),\label{trig-id2}\\
%&\tan(\al\pm\be)=\frac{\tan(\al)\pm \tan(\be)}{1\mp \tan(\al) \tan(\be)},\label{trig-id3}\\
&\sin(2\al)\pm \sin(2\be)=2\sin(\al\pm \be)\cos(\al\mp \be),\label{trig-id4}\\
&\cos(2\al)+\cos(2\be)=2\cos(\al+ \be)\cos(\al- \be),\label{trig-id5}\\
&\cos(2\al)-\cos(2\be)=-2\sin(\al+ \be)\sin(\al- \be),\label{trig-id6}\\
&\tan(\al)\pm \tan(\be)=\frac{\sin(\al\pm\be)}{\cos(\al)\cos(\be)},\label{trig-id7}\\
&\cot(\al)\pm \cot(\be)=\pm\frac{\sin(\al\pm\be)}{\sin(\al)\sin(\be)}.\label{trig-id8}
\end{align}
\end{subequations}
To connect them with discrete equations,
throughout this paper we assume
\begin{equation}\label{al-be-ga}
   \al=an+bm+\al_0,~~\beta=\t{\alpha},~~\gamma=\h{\alpha},
\end{equation}
i.e. $\beta$ and $\gamma$ are respectively one-step shift of $\alpha$ along $n$ and $m$-direction.
In the following, let us investigate these addition formulae case by case,
and show how these formulae paly roles in constructing quadrilateral equations and their solutions.

\subsection{\eqref{trig-id1}} \label{subsec-trig-id1}

Consider  formula \eqref{trig-id1} and its counterpart in $m$-direction,
\begin{subequations}\label{tri-sin-minus}
\begin{align}
    & \sin(\al-\be)=\sin(\al)\cos(\be)-\cos(\al)\sin(\be),\\
    & \sin(\al-\ga)=\sin(\al)\cos(\ga)-\cos(\al)\sin(\ga).
\end{align}
\end{subequations}
Defining
\begin{equation}\label{tran-sin-minus}
    u=\cos(\al),~~U=\sin(\al),~~ p=\sin(a),~~q=\sin(b),
\end{equation}
then \eqref{tri-sin-minus} is   written as
\begin{subequations}\label{lpmKdV-bt}
\begin{align}
    p=u\t U-U\t u,\\
    q=u\h U-U\h u,
\end{align}
\end{subequations}
which is considered as a more general coupled system that admits more solutions than  \eqref{tran-sin-minus}.
After checking compatibility $\t{\h u}=\h{\t u}$ one finds $U$ must satisfy equation
\begin{equation}
p(U\t{U}-\h{U}\h{\t{U}})-q(U\h{U}-\t{U}\h{\t{U}})=0,
\label{lpmKdV-U}
\end{equation}
which is known as the lattice potential modified Korteweg-de Vries (lpmKdV) equation.
Since $u$ and $U$ are symmetrically placed in \eqref{lpmKdV-bt}, the compatibility $\t{\h U}=\h{\t U}$ yields another lpmKdV equation
\begin{equation}
p(u\t{u}-\h{u}\h{\t{u}})-q(u\h{u}-\t{u}\h{\t{u}})=0.\label{lpmKdV-u}
\end{equation}
Thus, \eqref{lpmKdV-bt}, \eqref{lpmKdV-U} and \eqref{lpmKdV-u} compose a consistent triplet in which
\eqref{lpmKdV-bt} serves as an auto-BT for the  lpmKdV equation.
Obvious, \eqref{tran-sin-minus} provides trigonometric function solutions
and $p$ and $q$ are parameterised spacing parameters.
Note that transformation $U\to U+ku$ keeps  system \eqref{lpmKdV-bt} unchanged,
which means linear combinations of $u, U$ are still  solutions to the lpmKdV equation
provided $u, U$ satisfy \eqref{lpmKdV-bt} simultaneously.
In particularly when $u=\cos(\al),~U=\sin(\al)$ the combination
provides a scaling and phase shift for the previous solutions.

Next, let us introduce a variable $w$ by
\begin{equation}\label{w}
    w=U/u.
\end{equation}
Following that  \eqref{lpmKdV-bt} becomes
\begin{subequations}\label{lpmKdV-btt}
\begin{align}
    p=u\t u(\t w-w),\\
    q=u\h  u(\h w-w),
\end{align}
\end{subequations}
which is a deformation of \eqref{lpmKdV-bt}.
Consistent triplet resulting from \eqref{lpmKdV-bt} yields the lpmKdV equation \eqref{lpmKdV-u}
and
\begin{equation}\label{Q10}
    p^2(w-\h{w})(\t{w}-\h{\t{w}})-q^2(w-\t{w})(\h{w}-\h{\t{w}})=0,
\end{equation}
which is the cross-ratio equation (also known as Q1($0;p^2,q^2$) equation in the ABS list).
As a result,
$ w=\tan(\al)$
gives a solution to the above Q1(0) equation with $(p,q)$ parameterized in \eqref{tran-sin-minus}.

We can also get a Lax pair for the lpmKdV equation from its BT \eqref{lpmKdV-bt}. To achieve that,
introduce variable
\begin{equation}\label{z}
    z=uU,
\end{equation}
from which  \eqref{lpmKdV-bt} is written as
\begin{subequations}\label{lpmKdV-H31}
\begin{align}
    pu\t u=u^2\t z-\t u^2 z,\\
   qu\h u=u^2\h z-\h u^2 z.
\end{align}
\end{subequations}
It can  be linearized by introducing  $z=\frac{f}{\lambda g}$, i.e.
\begin{align*}
\frac{\t f}{\t g}=\frac{p^{-1}f\t u/u+\lambda g }{p^{-1}gu/\t u},~~
\frac{\h f}{\h g}=\frac{q^{-1}f\h u/u+\lambda g }{q^{-1}gu/\h u},
\end{align*}
which leads to a pair of linear problems
\begin{equation}\label{lax-pair}
    \t\phi=L\phi,~~\h\phi=M\phi
\end{equation}
where
\begin{equation*}
    \phi=\left(\begin{array}{c}
           f \\
           g
         \end{array}\right),~~
    L=\left(\begin{array}{cc}
p^{-1}\t u/u &  \lambda \\
 0&p^{-1}u/\t u \\
  \end{array}\right),~~
   M=\left( \begin{array}{cc}
q^{-1}\h u/u &  \lambda \\
 0&q^{-1}u/\h u \\
  \end{array}\right).
\end{equation*}
This is a Lax pair of  the lpmKdV equation \eqref{lpmKdV-u}, which is used in \cite{XC-MPLB-2016} to derive solutions for the equation.
Note that $\lambda$ is a fake spectral parameter as it can be removed by simple gauge transformation.

Besides, by direct calculation we can find system \eqref{lpmKdV-H31}
provides a  BT between the lpmKdV equation \eqref{lpmKdV-u} and  H3*($\de;P,Q$) equation (cf.\cite{AtkN-IMRN-2014}):
\begin{align}\label{H3*}
&(P-Q)\left[P(z\h{z}-\t{z}\h{\t{z}})^2-Q(z\t{z}-\h{z}\h{\t{z}})^2\right]\nonumber \\
&~~~ +(z-\h{\t{z}})(\t{z}-\h{z})\left[(z-\h{\t{z}})(\t{z}-\h{z})PQ-4\delta^2(P-Q)\right]=0,
\end{align}
where $P=4/p^2,~Q=4/q^2, \de=1$.
In addition, eliminating $U$ from \eqref{w} and \eqref{z}, we get
\[ u^2=z/w,\]
from which \eqref{lpmKdV-btt} yields
\begin{subequations}\label{Q10-H3*1}
\begin{align}
    p^2w\t w=z\t z(\t w-w)^2,\\
    q^2w\h w=z\h z(\h w-w)^2,
\end{align}
\end{subequations}
which gives a BT between  Q1($0;p^2,q^2$) and H3*($1;P,Q$) equation.

\subsection{\eqref{trig-id2}}\label{subsec-trig-id2}

Come to \eqref{trig-id2}. We consider
\begin{subequations}\label{tri-cos-minus}
\begin{align}
   & \cos(\alpha-\beta)=\cos(\al)\cos(\be)+\sin(\al)\sin(\be),\\
   & \cos(\alpha-\gamma)=\cos(\al)\cos(\ga)+\sin(\al)\sin(\ga),
\end{align}
\end{subequations}
and define
\begin{equation}\label{trans-2}
  u=\cos(\al),~~U=\sin(\al),~~ \de p=-\cos(a),~~\de q=-\cos(b).
\end{equation}
It then follows from \eqref{tri-cos-minus} that
\begin{subequations}\label{H3d-bt}
\begin{align}
    -\delta p=u\t u+U\t U,\\
      -\delta q=u\h u+U\h U,
\end{align}
\end{subequations}
which provides  an auto-BT for  H3($\delta$) equation,
\begin{subequations}\label{H3d-u-U}
\begin{align}
&   p(u\t{u}+\h{u}\h{\t{u}})-q(u\h{u}+\t{u}\h{\t{u}})+\delta(p^2-q^2)=0,\label{H3d-u}\\
&    p(U\t{U}+\h{U}\h{\t{U}})-q(U\h{U}+\t{U}\h{\t{U}})+\delta(p^2-q^2)=0.\label{H3d-U}
\end{align}
\end{subequations}
\eqref{H3d-bt} and  \eqref{H3d-u-U} constitutes a consistent triplet.

Moreover, substituting \eqref{w} into system \eqref{H3d-bt} yields
\begin{subequations}\label{H3d-w-bt}
\begin{align}
    -\delta p=u\t u(1+w\t w),\\
      -\delta q=u\h u(1+w\h w).
\end{align}
\end{subequations}
Compatibility of $w$ and $u$ respectively lead the H3($\delta$) equation \eqref{H3d-u} and equation
\begin{equation}\label{ww}
   p^2(1+w\h w)(1+\t w \h{\t w})=q^2(1+w\t w)(1+\h w \h{\t w}),
\end{equation}
which can be transformed to Q1($0$) equation \eqref{Q10} by $w\to (-1)^{n+m}w^{(-1)^{n+m}}$.
Consequently $w=\tan(\al)$ solves Eq.\eqref{ww} with $(p,q)$ parameterised as in  \eqref{trans-2}.

Again, substituting \eqref{z} into system \eqref{H3d-bt} leads to
\begin{subequations}\label{H3d-H3*-bt}
\begin{align}
    (\delta p+u\t u) u\t u=-z\t z,\\
     (\delta q+u\h u) u\h u=-z\h z,
\end{align}
\end{subequations}
which is a BT between H3($\delta$) equation \eqref{H3d-u} and H3*($\delta$) equation  \eqref{H3*}
with $P=-4/p^2,~Q=-4/q^2$.
Consequently $z=\sin(2\al)/2$ solves \eqref{H3*} with $ P=-4\de^2\sec^2(a),~Q=-4\de^2\sec^2(b)$.

Besides, relation between variables $z$ and $w$, which is
\begin{subequations}\label{w-H3*-bt}
\begin{align}
   & \delta^2 p^2 w\t w=z\t z(1+w\t w)^2,\\
   & \delta^2 q^2 w\h w=z\h z(1+w\h w)^2,
\end{align}
\end{subequations}
provides a BT between equation \eqref{ww} and \eqref{H3*} .

\subsection{\eqref{trig-id4}}\label{subsec-trig-id4}

\subsubsection{Case (I)}\label{sec-2-3-1}

Consider
\begin{subequations}\label{tri-sinplussin}
\begin{align}
  &  \sin(2\al)+\sin(2\be)=2\sin(\al+ \be)\cos(\al- \be),\\
  &  \sin(2\al)+\sin(2\ga)=2\sin(\al+ \ga)\cos(\al- \ga),
\end{align}
\end{subequations}
and define
\begin{equation}
 u=\sin(\al),~~U=\cos(\al), ~~p=\cos(a),~~q=\cos(b).
\label{uU-Q30}
\end{equation}
Making use of double-angle formula $\sin(2\al)=2\sin(\al)\cos(2\al)$,
system  \eqref{tri-sinplussin} yields
\begin{subequations}\label{Q30-bt}
\begin{align}
   &  uU+\t u \t U=p(u\t U+U\t u),\\
   &  uU+\h u \h U=q(u\h U+U\h u),
\end{align}
\end{subequations}
in which $u$ and $U$ appear symmetrically.
The compatibility $\t{\h{U}}=\h{\t{U}}$ leads to
\[(u\h{\t{u}}-\t u \h u)[p(1-q^2)(u\h{u}+\t{u}\h{\t{u}})-q(1-p^2)(u\t{u}+\h{u}\h{\t{u}})-(p^2-q^2)(\t{u}\h{u}+u\h{\t{u}})]=0.\]
Thus, if $u\h{\t{u}}\neq \t u \h u$,
\eqref{Q30-bt} provides an auto-BT for Q3(0) equation
\begin{equation}\label{Q30}
    p(1-q^2)(u\h{u}+\t{u}\h{\t{u}})-q(1-p^2)(u\t{u}+\h{u}\h{\t{u}})-(p^2-q^2)(\t{u}\h{u}+u\h{\t{u}})=0.
\end{equation}
As a result, both $u=\sin(\al)$ and $u=\cos(\al)$
solve  Q3(0) equation \eqref{Q30}.

Introducing $z$ by \eqref{z} and rewriting  \eqref{Q30-bt} in terms of $(u,z)$ as
\begin{subequations}\label{Q30-Q30*-bt}
\begin{align}
    p(u^2\t z+\t u^2 z)=u\t u(z+\t z),\\
   q(u^2\h z+\h u^2 z)=u\h u(z+\h z),
\end{align}
\end{subequations}
the compatibility of $u$ leads that $z$ satisfies
Q3*(0) equation (cf.\cite{AtkN-IMRN-2014})
\begin{align}\label{Q3*0}
&(P-Q)[P(z\t z-\h z\h{\t{z}})^2-Q(z\h z-\t z\h{\t{z}})^2]\nonumber\\
&~~~+(z-\h{\t{z}})(\t z-\h z)\left[(z-\h{\t{z}})(\t z-\h z)(PQ-1)-2(P-Q)(z\h{\t{z}}+\t z\h z)\right]=0,
\end{align}
with $P=2p^2-1, ~Q=2q^{2}-1$.
Thus \eqref{Q30-Q30*-bt} can be regarded as a BT for Q3(0) and Q3*(0) equation.
Then $z=\sin(2\al)/2$ solves Q3*(0) equation with $P=\cos(2a), Q=\cos(2b)$.

With \eqref{w} system \eqref{Q30-bt} is deformed as
\begin{subequations}\label{Q30-Q30*-bt1}
\begin{align}
    pu\t u(w+\t w)=u^2w+\t u^2\t w,\\
   qu\h u(w+\h w)=u^2w+\h u^2\h w.
\end{align}
\end{subequations}
One can check that this is a BT between Q3(0) equation \eqref{Q30} and Q3*(0) equation \eqref{Q3*0}
with parameters $P=2p^{-2}-1, ~Q=2q^{-2}-1$.
Hence both $w=\tan(\al)$ and $w=\cot(\al)$ are solutions of Q3*(0) equation \eqref{Q3*0} with $P=2\sec(a)^2-1, ~Q=2\sec(b)^2-1$.
In fact, it is also easy to see
\begin{equation}\label{trans-wz}
w\to z^{-1},~~p\to p^{-1}, ~~q\to q^{-1}
\end{equation}
brings  BT  \eqref{Q30-Q30*-bt1}  to BT \eqref{Q30-Q30*-bt}.

Similar to \eqref{Q10-H3*1}, 
from \eqref{Q30-Q30*-bt1}  and  \eqref{Q30-Q30*-bt} we can get
\begin{subequations}\label{Q30*-Q30*-bt}
\begin{align}
  &  p^2z\t z(w+\t w)^2=w\t w(z+\t z)^2,\\
  &  q^2z\h z(w+\h w)^2=w\h w(z+\h z)^2,
\end{align}
\end{subequations}
which is an auto-BT for Q3*(0) equation.

Using the relation between Q3(0) and A2 equation in the ABS list, the above results can be transformed to those of A2 equation.
These results are: with
\[  u=(\sin(\al))^{(-1)^{n+m}},~~U=(\cos(\al))^{(-1)^{n+m+1}}\]
system  \eqref{tri-sinplussin} indicates relation
\begin{subequations}\label{A2-bt}
\begin{align}
  &  u\t u+U \t U=p(1+u\t uU\t U),\\
  &  u\h u+U \h U=q(1+u\h uU\h U),
\end{align}
\end{subequations}
which turns out to be an  auto-BT for A2 equation
\begin{equation}\label{A2}
 p(1-q^2)(u\t{u}+\h{u}\h{\t{u}})-q(1-p^2)(u\h{u}+\t{u}\h{\t{u}})-(p^2-q^2)(1+u\t{u}\h{u}\h{\t{u}})=0,
\end{equation}
which admits solutions $u=(\sin(\al))^{(-1)^{n+m}}$ and $u=(\cos(\al))^{(-1)^{n+m}}$;
 with \eqref{z} system \eqref{A2-bt} is written as
\begin{subequations}\label{A2-A2*-bt1}
\begin{align}
    u\t u+\frac{z\t z}{u \t u}=p(1+z\t z),\\
   u\h u+\frac{z\h z}{u \h u}=q(1+z\h z),
\end{align}
\end{subequations}
which shows a connection between A2 equation \eqref{A2} and A2* equation (cf. \cite{AtkN-IMRN-2014})
\begin{align}\label{A2*}
& (P-Q)\left[P(z\h{z}-\t{z}\h{\t{z}})^2-Q(z\t{z}-\h{z}\h{\t{z}})^2\right] \nonumber\\
&~~~~+(z-\h{\t{z}})(\t{z}-\h{z}) \left[(z-\h{\t{z}})(\t{z}-\h{z})(PQ-1)+2(P-Q)(1+z\t{z}\h{z}\h{\t{z}})\right]=0,
\end{align}
where $P=2p^{-2}-1,~Q=2q^{-2}-1,$ and hence $z=(\tan(\al))^{(-1)^{n+m}}$ solves A2* equation with  $P=2\sec(a)^2-1, ~Q=2\sec(b)^2-1$;
similarly, through \eqref{w}, system \eqref{A2-bt} is deformed as
\begin{subequations}\label{A2-A2*-bt2}
\begin{align}
  &  u\t u(1+w \t w)=p(1+u^2\t u^2w\t w),\\
  &  u\h u(1+w \h w)=q(1+u^2\h u^2w\h w),
\end{align}
\end{subequations}
which is a BT for A2  and A2* equation
where $P=2p^2-1,~Q=2q^2-1,$
and therefore $w=(\sin(2\al)/2)^{(-1)^{n+m}}$ solves A2* equation with $P=\cos(2a),~Q=\cos(2b)$.
Also transformation \eqref{trans-wz} establishes the relation for BT  \eqref{A2-A2*-bt1} and \eqref{A2-A2*-bt2}.
In addition, system
\begin{subequations}\label{A2*-bt}
\begin{align}
  & z\t z(1+w\t w)^2=p^2 w\t w(1+z\t z)^2,\\
  & z\h z(1+w\h w)^2=q^2w\h w(1+z\h z)^2
\end{align}
\end{subequations}
gives an auto-BT for A2* equation.

\subsubsection{Case (II)}

For formulae
\begin{subequations}\label{tri-sin-sin}
\begin{align}
  &  \sin(2\al)-\sin(2\be)=2\sin(\al- \be)\cos(\al+ \be),\\
  &  \sin(2\al)-\sin(2\ga)=2\sin(\al- \ga)\cos(\al+ \ga),
\end{align}
\end{subequations}
we suppose
\[  u=\sin(\al),~~U=\cos(\al), ~~p=\sin(a),~~q=\sin(b).\]
It then gives
\begin{subequations}\label{lpmKdV-bt2}
\begin{align}
   & uU-\t u \t U=p(u\t u-U\t U),\\
   & uU-\h u \h U=q(u\h u-U\h U),
\end{align}
\end{subequations}
which is an auto-BT for the lpmKdV equation  \eqref{lpmKdV-U} and \eqref{lpmKdV-u}.

Via \eqref{z} the above system is deformed into
\begin{subequations}\label{lpmKdV-A2*-bt1}
\begin{align}
   & z-\t z=p(u\t u-\frac{z\t z}{u\t u}),\\
   & z-\h z=q(u\h u-\frac{z\h z}{u\h u}),
\end{align}
\end{subequations}
which is  a BT for lpmKdV equation \eqref{lpmKdV-u} and Q3*(0) equation \eqref{Q3*0} with $P=1-2p^2, Q=1-2q^2$;
while via \eqref{w}, system \eqref{lpmKdV-bt2} is written as
 \begin{subequations}\label{lpmKdV-A2*-bt2}
\begin{align}
   & u^2w-\t u^2\t w=pu\t u(1-w\t w),\\
   &  u^2w-\h u^2\h w=qu\h u(1-w\h w),
\end{align}
\end{subequations}
which turns out to be a BT for the lpmKdV equation \eqref{lpmKdV-u} and A2* equation \eqref{A2*} with $P=1-2p^{-2}, Q=1-2q^{-2}$.
The $(z,w)$ system obtained from \eqref{lpmKdV-A2*-bt1} and \eqref{lpmKdV-A2*-bt2},
 \begin{subequations}\label{A2*-Q3*0-bt}
\begin{align}
  & (z-\t z)^2w\t w=p^2z\t z(1-w\t w)^2,\\
  &  (z-\h z)^2w\h w=q^2z\h z(1-w\h w)^2,
\end{align}
\end{subequations}
provides a BT for A2* equation and  Q3*(0) equation.
As a result   $\sin(2\al)/2$ solves Q3*(0) equation  with $P=\cos(2a), Q=\cos(2b)$ and $\cot(\al)$ solves  A2* equation with
 $P=1-2\mathrm{csc}^2(a), Q=1-2\mathrm{csc}^2(b)$.

A second parametrisation  of \eqref{tri-sin-sin} is
\[  u=-\sin(2\al)/2,~~U=\tan(\al),~~p=\sin^2(a),~~q=\sin^2(b).\]
In this case \eqref{tri-sin-sin} yields
\begin{subequations}\label{Q11-Q10-bt}
\begin{align}
    \t u-u=\frac{p(U\t U-1)}{\t U-U},\\
    \h u-u=\frac{q(U\h U-1)}{\h U-U},
\end{align}
\end{subequations}
which acts as a BT for Q1($1$) equation
\begin{equation}\label{Q11}
     p(u-\h{u})(\t{u}-\h{\t{u}})-q(u-\t{u})(\h{u}-\h{\t{u}})+pq(p-q)=0
\end{equation}
and Q1(0) equation of $u$.

A third choice,
\[  u=\frac{1}{4}\left[\sin(2\al)-(\sin(2a)n+\sin(2b)m)\right],~~U=\tan(\al),~~p=\sin^2(a),~~q=\sin^2(b),\]
brings system \eqref{tri-sin-sin} to
\begin{subequations}\label{Q10-bt}
\begin{align}
  &  \t u-u=\frac{-pU\t U}{\t U-U},\\
  &  \h u-u=\frac{-qU\h U}{\h U-U},
\end{align}
\end{subequations}
which is an auto-BT for  Q1(0) equation.
Meanwhile, taking $w=u/U$  the BT \eqref{Q10-bt} leads to
\begin{subequations}\label{Q10--bt}
\begin{align}
    \t U\t w-Uw=\frac{-pU\t U}{\t U-U},\\
    \h U\h w-Uw=\frac{-qU\h U}{\h U-U},
\end{align}
\end{subequations}
which is a BT for Q1(0) equation of $U$ and Q2* equation
\begin{align}\label{Q2*}
   &(p-q)[p(w\t w-\h w\h{\t{w}})(w+\t w-\h w-\h{\t{w}})-q(w\h w-\t w\h{\t{w}})(w-\t w+\h w-\h{\t{w}})]\nonumber\\
&~~ +(w-\h{\t{w}})(\t w-\h w)\left[p(w-\h w)(\t w- \h{\t{w}})-q(w-\t w)(\h w- \h{\t{w}})-pq(p-q)\right]=0.
\end{align}
Hence
\[w= \frac{1}{4\tan(\al)}\left[\sin(2\al)-(\sin(2a)n+\sin(2b)m)\right]\]
solves  Q2* equation \eqref{Q2*}.

One more choice is
\[  u=-\frac{1}{4}\left[\sin(2\al)-(\sin(2a)n+\sin(2b)m)\right],~~U=\sin(\al),~~p=\sin(a),~q=\sin(b),\]
which brings \eqref{tri-sin-sin} to
\begin{subequations}\label{Q10-lpmkdv-bt}
\begin{align}
  &  \t u-u=pU\t U,\\
  &  \h u-u=qU\h U,
\end{align}
\end{subequations}
which gives a BT for Q1($0;p^2,q^2$) equation \eqref{Q10} of $u$
and lpmKdV equation \eqref{lpmKdV-U}.
%listed in the first entry of Table 1 in \cite{ZZ-arx-2017-1}.
Again considering $w=u/U$,
system \eqref{Q10-lpmkdv-bt} leads to
\begin{subequations} 
\begin{align}
  &  \t w  \t U-wU=pU\t U,\\
  &  \h w \h U-wU=qU\h U,
\end{align}
\end{subequations}
which gives an auto-BT for the lpmKdV equation. Then function
\[w=-\frac{1}{4\sin(\al)}\left[\sin(2\al)-(\sin(2a)n+\sin(2b)m)\right]\]
solves the lpmKdV equation.
Note that \eqref{Q10-lpmkdv-bt} played a pivot role in \cite{ZhaZ-SIGMA-2017} to
generate a series of rational solutions for some ABS equations.

\subsection{\eqref{trig-id5}}\label{subsec-trig-id5}
From formula \eqref{trig-id5}, we have
\begin{subequations}\label{tri-5}
\begin{align}
   & \cos(2\al)+\cos(2\be)=2\cos(\al+ \be)\cos(\al- \be),\\
   & \cos(2\al)+\cos(2\ga)=2\cos(\al+ \ga)\cos(\al- \ga).
\end{align}
\end{subequations}
Employing
\[
  u=\frac{\cos(2\al)}{-4\de},~~U=\cos(\al), ~~ \de p=-\cos(a),~~\de q=-\cos(b),
\]
system \eqref{tri-5} leads to
\begin{subequations}\label{A1d-bt}
\begin{align}
  &  u+\t u-\frac{\de}{2}p^2=pU\t U,\\
  &  u+\h u-\frac{\de}{2}q^2=qU\h U,
\end{align}
\end{subequations}
which can be regarded as a BT for
 H3($\delta$) equation \eqref{H3d-U} and
 A1($\delta$) equation
\[ P(u+\h{u})(\t{u}+\h{\t{u}})-Q(u+\t{u})(\h{u}+\h{\t{u}})-\delta^2PQ(P-Q)=0\]
where $P=p^2/2,~Q=q^2/2$.

In addition, from system \eqref{tri-5} and relation $ \cos(2\al)=2\cos^2(\al)-1$,
parametrisation
\[
  u=\cos(\al),~~U=\sin(\al),~~p=\cos(a),~~q=\cos(b)
\]
leads to
\begin{subequations}\label{cos-uU}
\begin{align}
   & u^2+\t u^2-1=p(u\t u-U \t U),\\
   &  u^2+\h u^2-1=q(u\h u-U \h U).
\end{align}
\end{subequations}
From the compatibility of $U$ in  \eqref{cos-uU}, we derive a  $u$   equation
\begin{equation}\label{multi-u}
    q^2(u^2+\t u^2-1-pu\t u)(\h u^2+\h{\t u}^2-1-p\h u\h{\t u})=p^2(u^2+\h u^2-1-qu\h u)(\t u^2+\h{\t u}^2-1-q\t u\h{\t u}),
\end{equation}
which is not included in the  multi-quadratic equations given in \cite{AtkN-IMRN-2014}.
Note that we do not have an explicit equation for $U$.

\subsection{\eqref{trig-id6}}\label{subsec-trig-id6}
Now we consider
\begin{subequations}\label{tri-6}
\begin{align}
  &  \cos(2\al)-\cos(2\be)=-2\sin(\al+ \be)\sin(\al- \be),\\
  &  \cos(2\al)-\cos(2\ga)=-2\sin(\al+ \ga)\sin(\al- \ga).
\end{align}
\end{subequations}
By
\[
  u=\cos(\al),~~U=\sin(\al),~~p=\sin(a),~~q=\sin(b),
\]
system \eqref{tri-6} yields
\begin{subequations}\label{lpmkdv-unknown-bt}
\begin{align}
  &  u^2-\t u^2=p(U\t u+u\t U),\\
  &  u^2 -\h u^2=q(U\h u+u\h U)
\end{align}
\end{subequations}
of which the compatibility $\t{\h{U}}=\h{\t{U}}$ leads to
\begin{equation}
(u\h{\t{u}}+\t u \h u)[p(u\t u-\h u \h{\t u})-q(u\h u-\t u \h{\t u})]=0,
\end{equation}
which is the lpmKdV equation multiplied by the term $u\h{\t{u}}+\t u \h u$.
Since system \eqref{lpmkdv-unknown-bt} is linear in terms of $U$, it can be viewed as a weak Lax pair (cf.\cite{HieV-Non-2011}) for the lpmKdV equation.
$u=\cos(\al)$ provides a solution for the lpmKdV equation as $u\h{\t{u}}+\t u \h u\neq 0$.
Note that we can not eliminate variable $u$ to have a neat form for equation of $U$.

Besides,  \eqref{lpmkdv-unknown-bt} with \eqref{w} is written as
\begin{subequations}\label{lpmkdv-A1*-bt}
\begin{align}
  &  u^2-\t u^2=pu\t u(w+\t w),\\
  &  u^2 -\h u^2=pu\h u(w+\h w),
\end{align}
\end{subequations}
which is a BT connecting solutions between the lpmKdV equation \eqref{lpmKdV-u} and A1* equation
\begin{align}\label{A1*}
&(P-Q)\left[P(w-\t w+\h w-\h{\t w})^2-Q(w-\h w+\t w-\h{\t w})^2\right]\nonumber \\
&~~ +4(w-\h{\t w})(\t w-\h w)\left[P(w+\h w)(\t w+\h{\t w})-Q(w+\t w)(\h w+\h{\t w})\right]=0,
\end{align}
where $P=-4 p^{-2},~~Q=-4 q^{-2}$.
Hence we have a solution $w=\tan(\al)$ for A1* equation \eqref{A1*} with $P=-4\csc^2(a),~Q=-4\csc^2(b)$.
Further, with
\[ u=(\mathrm{i}^{n+m} v)^{(-1)^{n+m}},~~ p=2\mathrm{i}p'^{-1},~~ q=2\mathrm{i}q'^{-1},~~\mathrm{i}^2=-1 \]
BT \eqref{lpmkdv-A1*-bt} turns out  to be
\begin{subequations}\label{H30-A1*-bt}
\begin{align}
 &   (v^2\t v^2+1)p'=2v\t v(w+\t w),\\
 &  (v^2\h v^2+1)q'=2v\h v(w+\h w)
\end{align}
\end{subequations}
which is a BT for H3$(0;p',q')$ equation
\begin{equation}\label{H30}
    p'(v\t{v}+\h{v}\h{\t{v}})-q'(v\h{v}+\t{v}\h{\t{v}})=0
\end{equation}
and A1* equation \eqref{A1*}.

\subsection{\eqref{trig-id7} (or \eqref{trig-id8})}\label{subsec-trig-id7}

As for formula \eqref{trig-id7} (or \eqref{trig-id8}) and its $(\gamma,\h{~~})$ counterpart, it is interesting to observe that parametrisation
\[ u=\tan(\al), ~~U=\sec(\al),~~p=\sin(a),~~q=\sin(b)\]
brings us system \eqref{Q10-lpmkdv-bt},
which means we can then share those results related to \eqref{Q10-lpmkdv-bt}.

\section{Cases of elliptic functions}\label{sec-3}

\subsection{Jacobi elliptic functions}\label{sec-3-1}

Jacobi elliptic function $s=\mathrm{sn}(z,k)$   is defined by  elliptic integration
\[   z=\int^s_0\frac{dt}{\sqrt{(1-t^2)(1-k^2t^2)}}\]
where the parameter $k$ is called  modulus, and we note that $s$ satisfies the first order differential equation
\[ {s'}^2=(1-s^2)(1-k^2s^2)\]
where $s'=\frac{\partial s}{\partial z}$. Taking a limit $k\to 0$, the sn-function reduces to the sine function.
Jacobi's sn, cn and dn-function are  related by
\begin{eqnarray*}
&&\mathrm{cn}^2(z)+\mathrm{sn}^2(z)=1,\\
&&\mathrm{dn}^2(z)+k^2\mathrm{sn}^2(z)=1,\\
&&\mathrm{sn}'(z)=\mathrm{cn}(z)\mathrm{dn}(z),
\end{eqnarray*}
where we have dropped off $k$ without any confusion.

Consider
\begin{subequations}\label{formula-sn}
\begin{align}
   & \mathrm{sn}(\al-\be)=\frac{\mathrm{sn}^2(\al)-\mathrm{sn}^2(\be)}
    {\mathrm{sn}(\al)\mathrm{cn}(\be)\mathrm{dn}(\be)+\mathrm{sn}(\be)\mathrm{cn}(\al)\mathrm{dn}(\al)},\\
   & \mathrm{sn}(\al-\ga)=\frac{\mathrm{sn}^2(\al)-\mathrm{sn}^2(\ga)}
    {\mathrm{sn}(\al)\mathrm{cn}(\ga)\mathrm{dn}(\ga)+\mathrm{sn}(\ga)\mathrm{cn}(\al)\mathrm{dn}(\al)}.
\end{align}
\end{subequations}
Taking
\[u=\mathrm{sn}(\al),~~U=\mathrm{cn}(\al)\mathrm{dn}(\al),~~p=\mathrm{sn}(a),~~q=\mathrm{sn}(b),\]
from \eqref{formula-sn} we can obtain system \eqref{lpmkdv-unknown-bt}
and further via \eqref{w} we obtain  \eqref{lpmkdv-A1*-bt}.
Consequently function $u=\mathrm{sn}(\al)$ provides a solution to the lpmKdV equation \eqref{lpmKdV-U}
and $w=\frac{\mathrm{cn}(\al)\mathrm{dn}(\al)}{\mathrm{sn}(\al)}$
solves A1* equation \eqref{A1*} with parameters
$P=-4\mathrm{sn}^{-2}(a),~~Q=-4\mathrm{sn}^{-2}(b)$.

Next, let us consider
\begin{subequations}\label{formula-cn}
\begin{align}
  &  \mathrm{cn}(\al-\be)=\frac{\mathrm{sn}(\al)\mathrm{cn}(\al)\mathrm{dn}(\be)+\mathrm{sn}(\be)\mathrm{cn}(\be)\mathrm{dn}(\al)}
    {\mathrm{sn}(\al)\mathrm{cn}(\be)\mathrm{dn}(\be)+\mathrm{sn}(\be)\mathrm{cn}(\al)\mathrm{dn}(\al)},\\
  &  \mathrm{cn}(\al-\ga)=\frac{\mathrm{sn}(\al)\mathrm{cn}(\al)\mathrm{dn}(\ga)+\mathrm{sn}(\ga)\mathrm{cn}(\ga)\mathrm{dn}(\al)}
    {\mathrm{sn}(\al)\mathrm{cn}(\ga)\mathrm{dn}(\ga)+\mathrm{sn}(\ga)\mathrm{cn}(\al)\mathrm{dn}(\al)}.
\end{align}
\end{subequations}
With
\[u=\mathrm{cn}(\al),~~U=\frac{\mathrm{sn}(\al)}{\mathrm{dn}(\al)},~~p=\mathrm{cn}(a),~~q=\mathrm{cn}(b)\]
\eqref{formula-cn} yields  the system \eqref{Q30-bt} in Sec.\ref{sec-2-3-1},
Thus we can have results  parallel to Sec.\ref{sec-2-3-1}.
which will be listed in Table 2.

Now we come to
\begin{subequations}\label{formula-dn}
\begin{align}
  &  \mathrm{dn}(\al-\be)=\frac{\mathrm{sn}(\al)\mathrm{cn}(\be)\mathrm{dn}(\al)+\mathrm{sn}(\be)\mathrm{cn}(\al)\mathrm{dn}(\be)}
    {\mathrm{sn}(\al)\mathrm{cn}(\be)\mathrm{dn}(\be)+\mathrm{sn}(\be)\mathrm{cn}(\al)\mathrm{dn}(\al)},\\
  &  \mathrm{dn}(\al-\ga)=\frac{\mathrm{sn}(\al)\mathrm{cn}(\ga)\mathrm{dn}(\al)+\mathrm{sn}(\ga)\mathrm{cn}(\al)\mathrm{dn}(\ga)}
    {\mathrm{sn}(\al)\mathrm{cn}(\ga)\mathrm{dn}(\ga)+\mathrm{sn}(\ga)\mathrm{cn}(\al)\mathrm{dn}(\al)}.
\end{align}
\end{subequations}
With
\[u=\mathrm{dn}(\al),~~U=\frac{\mathrm{sn}(\al)}{\mathrm{cn}(\al)},~~p=\mathrm{dn}(a),~~q=\mathrm{dn}(b), \]
\eqref{formula-dn} is converted to \eqref{Q30-bt} as well,
which means \eqref{formula-dn}  also leads to the results  parallel to Sec.\ref{sec-2-3-1}.
Again, we leave these results in Table 2.

\subsection{Weierstrass $\wp$ function}\label{sec-3-1}

Addition formula of the  Weierstrass elliptic $\wp$ function is
\begin{subequations}\label{formula-p}
\begin{align}
& \wp(\al)+\wp(\be)+\wp(\al-\be)=\frac{1}{4}\left(\frac{\wp'(\al)+\wp'(\be)}{\wp(\al)-\wp(\be)}\right)^2,\\
& \wp(\al)+\wp(\ga)+\wp(\al-\ga)=\frac{1}{4}\left(\frac{\wp'(\al)+\wp'(\ga)}{\wp(\al)-\wp(\ga)}\right)^2.
\end{align}
\end{subequations}
Introducing
\begin{equation}
u=\wp(\al)£¬~~U=\wp'(\al),~~p=\wp(a),~~q=\wp(b),
\end{equation}
system \eqref{formula-p} yields
\begin{subequations}\label{unknown-33}
\begin{align}
& u+\t u +p=\frac{1}{4}\left(\frac{U+\t U}{u-\t u}\right)^2,\\
& u+\h u +q=\frac{1}{4}\left(\frac{U+\h U}{u-\h u}\right)^2,
\end{align}
\end{subequations}
from which eliminating $U$  we find $u=\wp(\al)$ satisfies H2* equation (cf.\cite{AtkN-IMRN-2014})
\begin{align}\label{H2*}
&(p-q)\left[p(u-\t u+\h u-\h{\t u})^2-q(u-\h u+\t u-\h{\t u})^2\right]\nonumber \\
&~~~~+(u-\h{\t u})(\t u-\h u)\left[(u-\h{\t u})(\t u-\h u)-2(p-q)(u+\t u+\h u+\h{\t u})\right]=0.
\end{align}

Another verion of  addition formula of $\wp$ function presented in terms of Weierstrass  $\zeta$ function is
\begin{subequations}\label{formula-pz}
\begin{align}
& \wp(\al)+\wp(\be)+\wp(\al-\be)=(\zeta(\alpha)-\zeta(\beta)-\zeta(\alpha-\beta))^2,\\
& \wp(\al)+\wp(\ga)+\wp(\al-\ga)=(\zeta(\alpha)-\zeta(\ga)-\zeta(\alpha-\ga))^2.
\end{align}
\end{subequations}
In this turn we introduce
\begin{equation}
u=\wp(\al), ~~U=\zeta(\al)-n\zeta(a)-m\zeta(b),~~p=\wp(a),~~q=\wp(b),
\label{uU}
\end{equation}
and system \eqref{formula-p} yields
\begin{subequations}\label{unknown-33}
\begin{align}
& u+\t u +p=(U-\t U)^2,\\
& u+\h u +q=(U-\h U)^2,
\end{align}
\end{subequations}
which provides a BT between H2* equation \eqref{H2*} and H1 equation (i.e. the lpKdV)
\begin{equation}
(U-\th U)(\h U-\t U)=p-q.
\label{H1}
\end{equation}
Note that $U$ in \eqref{uU} was first presented in \cite{NijA-IMRN-2010} as a background solution of the lpKdV equation.

\section{Conclusion and discussion}\label{sec-4}

In this paper we have shown that addition formulae of many trigonometric functions and elliptic functions
are extended to 2-component systems of the form \eqref{BT-f};
considering solvability of \eqref{BT-f} one can derive a quadrilateral equation \eqref{u}
which appears as a compatibility of $U$ in \eqref{BT-f}
and a quadrilateral equation \eqref{U} which appears as a compatibility of $u$ in \eqref{BT-f}.
In this sense \eqref{BT-f}, \eqref{u} and \eqref{U} compose  a consistent triplet
in which \eqref{BT-f} act as a BT to connect two equations \eqref{u} and \eqref{U}.
Once a BT is obtained, it seems that the tricks of introducing $w$ by \eqref{w} and $z$ by \eqref{z}
works in deriving more BTs.
Most of the BTs obtained from addition formulae \eqref{trig-id} in this paper are already found in
\cite{Atk-JPA-2008,AtkN-IMRN-2014,ZhaZ-SIGMA-2017} from other approaches,
which means these addition formulae of trigonometric functions
reveal more general relations than the formulae themselves.
The reason behind this fact might be the following:
$\sin$ and $\cos$ functions can be viewed as alternative
forms of discrete plain wave factors which are usually in the form
$\bigl(\frac{a+k}{a-k}\bigr)^n\bigl(\frac{b+k}{b-k}\bigr)^m$
for quadrilateral equations.

We list out all trigonometric solutions we obtained, together with the related BTs, in Table 1,
and  elliptic solutions in Table 2.  In these tables spacing parameters $P_i$ and $Q_i$ are defined as the following,
\begin{align*}
&P_1=p^2,~~Q_1=q^2,\\
&P_2=4p^{-2},~~Q_2=4q^{-2},\\
&P_3=2p^{2}-1,~~Q_3=2q^{2}-1,\\
&P_4=2p^{-2}-1,~~Q_4=2q^{-2}-1.
\end{align*}
For those equations in which spacing parameters are not indicated, we mean they are $p$ and $q$.
Some solutions are same due to properties of trigonometric and elliptic functions.
For example, replacing $\al_0$ in  \eqref{al-be-ga} by $\al_0+\frac{\pi}{2}$,
then $\sin(\al)$ is replaced by $\cos(\al)$.
One more, replacing $\al_0$ in  \eqref{al-be-ga} by $\al_0+K$ where
$K=\int^1_0\frac{\mathrm{d}t}{\sqrt{1-t^2}\sqrt{1-k^2t^2}}$,
then $\mathrm{sn}(\al)$ can be replaced by $\frac{\mathrm{cn}(\alpha)}{\mathrm{dn}(\alpha)}$,
which means if $\mathrm{sn}(\al)$ solves the lpmKdV equation, so does $\frac{\mathrm{cn}(\alpha)}{\mathrm{dn}(\alpha)}$.
Note also that trigonometric solutions can be converted to nonperiodic ones expressed in terms of hyperbolic functions.

One can also define a $(N,M)$-periodic quadrilateral equation. For example, the lpmKdV equation \eqref{lpmKdV-u}
with $u(n,m)=u(n+lN,m+kM)$, ($l,k\in \mathbb{Z}$)
and $p=\sin(\al),~ q=\cos(\al)$.
The equation has a solution $u=\sin(\al)$ where $a=\frac{2\pi}{N},~b=\frac{2\pi}{M}$.
The solution $u=\mathrm{sn}(\al)$ enables one to consider the lpmKdV equation on a complex plane with double complex periods.
Besides, we are also interested in what soliton solutions and high order periodic solutions
are connected to these periodic solutions as background (seed) solutions.
This will be investigated in the future research.

\begin{landscape}
{\small\begin{center}
\begin{center}
\setlength{\tabcolsep}{12pt}
\renewcommand{\arraystretch}{1.5}
\small
\begin{tabular}{llllll}
\hline
   No.& \qquad   BT      &        $u$-equation  &   $U$-equation & parametrisation & ~~~~~~solutions\\
\hline
1&$ u\t U-U\t u=p$& lpmKdV &  lpmKdV &$p=\sin(a),~~q=\sin(b)$& $u=\cos(\al),~~U=\sin(\al)$\\
2&$ u\t u(\t U-U)=p $& lpmKdV & Q1(0;$P_1,Q_1$) & $p=\sin(a),~~q=\sin(b)$ & $u=\cos(\al),~~U=\tan(\al)$\\
3&$u^2\t U-\t u^2U=pu\t u$& lpmKdV &H3*($\de=1;P_2,Q_2$) &$p=\sin(a),~~q=\sin(b)$&$u=\cos(\al),~~U=\sin(2\al)/2$\\
4&$u\t u+U\t U=-\delta p$& H3($\de$) &H3($\de$) &$\de p=-\cos(a)$,~$\de q=-\cos(b)$  & $u=\cos(\al),~~U=\sin(\al)$\\
% &                       &           &          &$\de q=-\cos(b)$ & \\
5&$u\t u(1+U\t U)=-\delta p$& H3($\delta$) & \eqref{ww} &$\de p=-\cos(a),~\de q=-\cos(b)$ &$u=\cos(\al),~~U=\tan(\al)$ \\
6&$ (\delta p+u\t u) u\t u=-U\t U$&  H3($\delta$) &    H3*($\delta;-P_2,-Q_2$)  &$\de p=-\cos(a),~\de q=-\cos(b)$&$u=\cos(\al),~~U=\sin(2\al)/2$ \\
7&$U\t U(1+u\t u)^2=\de^2p^2u\t u$&  \eqref{ww} &    H3*($\delta;-P_2,-Q_2$)  &$\de p=-\cos(a),~\de q=-\cos(b)$&$u=\tan(\al),~~U=\sin(2\al)/2$ \\
8&$uU+\t u \t U=p(u\t U+U\t u)$& Q3($0$) &  Q3($0$)&$p=\cos(a),~~q=\cos(b)$&$u=\sin(\al),~~U=\cos(\al)$\\
9&$ u\t u(U+\t U)=p(u^2\t U+\t u^2U)$& Q3($0$) &  Q3*($0;P_3,Q_3$)& $p=\cos(a),~~q=\cos(b)$&$u=\sin(\al),~~U=\sin(2\al)/2$\\
10&$ pu\t u(U+\t U)=u^2U+\t u^2\t U$& Q3($0$) &  Q3*($0;P_4,Q_4$)& $p=\cos(a),~~q=\cos(b)$&$u=\sin(\al),~~U=\cot(\al)$\\
11&$p^2u\t u(U+\t U)^2=U\t U(u+\t u)^2$&  Q3*($0;P_3,Q_3$) &  Q3*($0;P_4,Q_4$)& $p=\cos(a),~~q=\cos(b)$&$u=\sin(2\al)/2,~~U=\cot(\al)$\\
12&$u\t u+U \t U=p(1+u\t uU\t U)$& A2 & A2 &$p=\cos(a),~~q=\cos(b)$&$ u=(\sin(\al))^{(-1)^{n+m}}$ \\
&&&&&$U=(\cos(\al))^{(-1)^{n+m+1}}$\\
13&$u\t u+\frac{U \t U}{u\t u}=p(1+U\t U)$& A2 & A2*($P_4,Q_4$) &$p=\cos(a),~~q=\cos(b)$&$ u=(\sin(\al))^{(-1)^{n+m}}$ \\
&&&&&$U=(\tan(\al))^{(-1)^{n+m}}$\\
14&$ u\t u(1+U \t U)=p(1+u^2\t u^2U\t U)$& A2 &  A2*($P_3,Q_3$)&$p=\cos(a),~~q=\cos(b)$ &$u=(\sin(\al))^{(-1)^{n+m}}$\\
&&&&&$U=(\sin(2\al)/2)^{(-1)^{n+m+1}}$\\
15&$ u\t u(1+U \t U)^2=p^2U\t U(1+u\t u)^2$& A2*($P_4,Q_4$) &  A2*($P_3,Q_3$)&$p=\cos(a),~~q=\cos(b)$ &$u=(\tan(\al))^{(-1)^{n+m}}$\\
&&&&&$U=(\sin(2\al)/2)^{(-1)^{n+m+1}}$\\
\hline
\end{tabular}
\begin{center}{\scriptsize {\bf Table 1.}   BTs with trigonometric functions.}
\end{center}
\end{center}
\end{center}
}
\end{landscape}

\begin{landscape}
{\small\begin{center}
\begin{center}
\setlength{\tabcolsep}{12pt}
\renewcommand{\arraystretch}{1.5}
\begin{tabular}{llllll}
\hline
   No.& \qquad   BT      &        $u$-equation  &   $U$-equation & parametrisation & ~~~~~~solutions\\
\hline
16&$ uU-\t u \t U=p(u\t u-U\t U)$&  lpmKdV &  lpmKdV &$p=\sin(a),~~q=\sin(b)$&$u=\sin(\al),~~U=\cos(\al)$\\
17&$ U-\t U=p(u\t u-\frac{U\t U}{u\t u})$&  lpmKdV &  Q3*(0;$-P_3,-Q_3$) &$p=\sin(a),~~q=\sin(b)$&$u=\sin(\al),~~U=\sin(2\al)/2$\\
18&$ u^2U-\t u^2\t U=pu\t u(1-U\t U)$&  lpmKdV &  A2*($-P_4,-Q_4$) &$p=\sin(a),~~q=\sin(b)$&$u=\sin(\al),~~U=\cot(2\al)$\\
19&$ (u-\t u)^2U\t U=p^2u\t u(1-U\t U)^2$&  Q3*(0;$-P_3,-Q_3$) &  A2*($-P_4,-Q_4$) &$p=\sin(a),~q=\sin(b)$&$u=\sin(2\al)/2,~~U=\cot(2\al)$\\
20&$ \t u-u=\frac{p(U\t U-1)}{\t U-U} $& Q1(1) &  Q1(0) & $p=\sin^2(a),~~q=\sin^2(b)$&$
u=-\sin(2\al)/2,~~U=\tan(\al)$\\
21&$  \t u-u=\frac{-pU\t U}{\t U-U} $& Q1(0) &  Q1(0) & $p=\sin^2(a),~~q=\sin^2(b)$&$ u=\frac{\sin(2\al)-(\sin(2a)n+\sin(2b)m)}{4},$\\
&&&&&$U=\tan(\al)$\\
22&$ \t U\t u-Uu=\frac{-pU\t U}{\t U-U} $& Q2* &   Q1(0) &$p=\sin^2(a),~~q=\sin^2(b)$ &$ u=\frac{\sin(2\al)-(\sin(2a)n+\sin(2b)m)}{4\tan(\al)},$\\
&&&&&$U=\tan(\al)$\\
23&$ \t u-u=pU\t U $&  Q1($0;P_1,Q_1$) &  lpmKdV &$p=\sin(a),~~q=\sin(b)$&$ u=-\frac{\sin(2\al)-\sin(2a)n-\sin(2b)m}{4},$\\
&&&&&$U=\sin(\al)$\\
&&&&&$u=\tan(\al), ~~U=\sec(\al)$\\
24&$  \t u  \t U-uU=pU\t U $& lpmKdV &  lpmKdV & $p=\sin(a),~~q=\sin(b)$&$ u=-\frac{\sin(2\al)-\sin(2a)n-\sin(2b)m}{4\sin(\al)},$\\
&&&&&$U=\sin(\al)$\\
25&$ u+\t u-\frac{\de}{2}p^2=pU\t U $& A1($\delta;P_1/2,Q_1/2$) &  H3($\delta$) &$ \de p=-\cos(a),$&$ u=\frac{\cos(2\al)}{-4\de},~U=\cos(\al)$\\
&&&& $\de q=-\cos(b)$ &\\
26&$  u^2-\t u^2=pu\t u(U+\t U) $& lpmKdV &  A1*($\delta;-P_2,-Q_2$) &$ p=\sin(a),~~q=\sin(b)$&$ u=\cos(\al),~~U=\tan(\al)$\\
\hline
\end{tabular}
\begin{center}{\scriptsize {\bf Table 1.}   Continue.}
\end{center}
\end{center}
\end{center}
}
\end{landscape}

\begin{landscape}
{\small\begin{center}
\begin{center}
\setlength{\tabcolsep}{12pt}
\renewcommand{\arraystretch}{1.5}
\begin{tabular}{llllll}
\hline
   No.& \qquad   BT      &        $u$-equation  &   $U$-equation & parametrisation & ~~~~~~solutions\\
\hline
1&$uU+\t u \t U=p(u\t U+U\t u)$& Q3($0$) &  Q3($0$)&$p=\mathrm{cn}(a),~~q=\mathrm{cn}(b)$&$u=\mathrm{cn}(\al),~~U=\frac{\mathrm{sn}(\al)}{\mathrm{dn}(\al)}$\\
&&&&$p=\mathrm{dn}(a),~~q=\mathrm{dn}(b)$&$u=\mathrm{dn}(\al),~~U=\frac{\mathrm{sn}(\al)}{\mathrm{cn}(\al)}$\\
2&$ u\t u(U+\t U)=p(u^2\t U+\t u^2U)$& Q3($0$) &  Q3*($0;P_3,Q_3$)&$p=\mathrm{cn}(a),~~q=\mathrm{cn}(b)$
&$u=\mathrm{cn}(\al),~~U=\frac{\mathrm{cn}(\al)\mathrm{sn}(\al)}{\mathrm{dn}(\al)}$\\
&&&&$p=\mathrm{dn}(a),~~q=\mathrm{dn}(b)$&$u=\mathrm{dn}(\al),~~U=\frac{\mathrm{sn}(\al)\mathrm{dn}(\al)}{\mathrm{cn}(\al)}$\\
3&$ pu\t u(U+\t U)=u^2U+\t u^2\t U$& Q3($0$) &  Q3*($0;P_4,Q_4$)&$p=\mathrm{cn}(a),~~q=\mathrm{cn}(b)$
&$u=\mathrm{cn}(\al),~~U=\frac{\mathrm{sn}(\al)}{\mathrm{dn}(\al)\mathrm{cn}(\al)}$\\
&&&&$p=\mathrm{dn}(a),~~q=\mathrm{dn}(b)$&$u=\mathrm{dn}(\al),~~U=\frac{\mathrm{sn}(\al)}{\mathrm{cn}(\al)\mathrm{dn}(\al)}$\\
4&$p^2u\t u(U+\t U)^2=U\t U(u+\t u)^2$&  Q3*($0;P_3,Q_3$) &  Q3*($0;P_4,Q_4$)&$p=\mathrm{cn}(a),~~q=\mathrm{cn}(b)$
&$u=\frac{\mathrm{cn}(\al)\mathrm{sn}(\al)}{\mathrm{dn}(\al)},~~U=\frac{\mathrm{sn}(\al)}{\mathrm{dn}(\al)\mathrm{cn}(\al)}$\\
&&&&$p=\mathrm{dn}(a),~~q=\mathrm{dn}(b)$
&$u=\frac{\mathrm{sn}(\al)\mathrm{dn}(\al)}{\mathrm{cn}(\al)},~~U=\frac{\mathrm{sn}(\al)}{\mathrm{cn}(\al)\mathrm{dn}(\al)}$\\
5&$u\t u+U \t U=p(1+u\t uU\t U)$& A2 & A2 &$p=\mathrm{cn}(a),~~q=\mathrm{cn}(b)$&$u=(\mathrm{cn}(\al))^{(-1)^{n+m}},$\\
&&&&&$U=\left(\frac{\mathrm{sn}(\al)}{\mathrm{dn}(\al)}\right)^{(-1)^{n+m+1}}$\\
&&&&$p=\mathrm{dn}(a),~~q=\mathrm{dn}(b)$&$u=(\mathrm{dn}(\al))^{(-1)^{n+m}},$\\
&&&&&$U=\left(\frac{\mathrm{sn}(\al)}{\mathrm{cn}(\al)}\right)^{(-1)^{n+m+1}}$\\
6&$u\t u+\frac{U \t U}{u\t u}=p(1+U\t U)$& A2 & A2*($P_4,Q_4$) &$p=\mathrm{cn}(a),~~q=\mathrm{cn}(b)$&$u=(\mathrm{cn}(\al))^{(-1)^{n+m}},$\\
&&&&&$U=\left(\frac{\mathrm{cn}(\al)\mathrm{dn}(\al)}{\mathrm{sn}(\al)}\right)^{(-1)^{n+m}}$\\
&&&&$p=\mathrm{dn}(a),~~q=\mathrm{dn}(b)$&$u=(\mathrm{dn}(\al))^{(-1)^{n+m}},$\\
&&&&&$U=\left(\frac{\mathrm{cn}(\al)\mathrm{dn}(\al)}{\mathrm{sn}(\al)}\right)^{(-1)^{n+m}}$\\
\hline
\end{tabular}
\begin{center}{\scriptsize {\bf Table 2.}   BTs with elliptic functions.}
\end{center}
\end{center}
\end{center}
}
\end{landscape}

\begin{landscape}
{\small\begin{center}
\begin{center}
\setlength{\tabcolsep}{12pt}
\renewcommand{\arraystretch}{1.5}
\begin{tabular}{llllll}
\hline
   No.& \qquad   BT      &        $u$-equation  &   $U$-equation & parametrisation & ~~~~~~solutions\\
\hline
7&$ u\t u(1+U \t U)=p(1+u^2\t u^2U\t U)$& A2 &  A2*($P_3,Q_3$)&$p=\mathrm{cn}(a),~q=\mathrm{cn}(b)$&$u=(\mathrm{cn}(\al))^{(-1)^{n+m}},$\\
&&&&&$U=\left(\frac{\mathrm{sn}(\al)\mathrm{cn}(\al)}{\mathrm{dn}(\al)}\right)^{(-1)^{n+m+1}}$\\
&&&&$p=\mathrm{dn}(a),~q=\mathrm{dn}(b)$&$u=(\mathrm{dn}(\al))^{(-1)^{n+m}},$\\
&&&&&$U=\left(\frac{\mathrm{sn}(\al)\mathrm{dn}(\al)}{\mathrm{cn}(\al)}\right)^{(-1)^{n+m+1}}$\\
8&$ u\t u(1+U \t U)^2=p^2U\t U(1+u\t u)^2$& A2*($P_4,Q_4$) &  A2*($P_3,Q_3$)&$p=\mathrm{cn}(a),~q=\mathrm{cn}(b)$
&$u=\left(\frac{\mathrm{cn}(\al)\mathrm{dn}(\al)}{\mathrm{sn}(\al)}\right)^{(-1)^{n+m}},$\\
&&&&&$U=\left(\frac{\mathrm{sn}(\al)\mathrm{cn}(\al)}{\mathrm{dn}(\al)}\right)^{(-1)^{n+m+1}}$\\
&&&&$p=\mathrm{dn}(a),~q=\mathrm{dn}(b)$&$u=\left(\frac{\mathrm{cn}(\al)\mathrm{dn}(\al)}{\mathrm{sn}(\al)}\right)^{(-1)^{n+m}},$\\
&&&&&$U=\left(\frac{\mathrm{sn}(\al)\mathrm{dn}(\al)}{\mathrm{cn}(\al)}\right)^{(-1)^{n+m+1}}$\\
9&$u^2-\t u^2=pu\t u(U+\t U) $& lpmKdV &  A1*($\delta;-P_2,-Q_2$) &$p=\mathrm{sn}(a),~q=\mathrm{sn}(b)$
&$u=\mathrm{sn}(\al),~U=\frac{\mathrm{cn}(\al)\mathrm{dn}(\al)}{\mathrm{sn}(\al)}$\\
10&$ u+\t u +p=(U-\t U)^2 $& H2* &  H1($-p,-q$) &$p=\wp(a),~q=\wp(b)$&$u=\wp(\al),~U=\zeta(\al)-n\zeta(a)-m\zeta(b)$\\
\hline
\end{tabular}
\begin{center}{\scriptsize {\bf Table 2.}   Continue.}
\end{center}
\end{center}
\end{center}
}
\end{landscape}

\vskip 15pt
\subsection*{Acknowledgments}
This project is  supported by the NSF of China (Nos.11371241 and 11631007).

\end{document}